\documentclass[
reprint,
floatfix,
aps,
prb,
amsmath,
twocolumn,
superscriptaddress,
amssymb,
tightenlines,
]{revtex4-1}

\pdfoutput=1
\usepackage{amsfonts,amssymb}
\usepackage[subnum]{cases}
\usepackage{mathrsfs}
\usepackage{amsmath}
\usepackage[none]{hyphenat}
\usepackage{hyperref} % links
\usepackage{bm} % bold math letters
\usepackage{natbib}
\usepackage{soul} % in order to highlight
\usepackage{color}
\usepackage{longtable}
\usepackage{ textcomp }
\usepackage{verbatim}
\usepackage{fancyhdr} % Required for custom headers
\usepackage{lastpage} % Required to determine the last page for the footer
\usepackage{extramarks} % Required for headers and footers
\usepackage{graphicx} % Required to insert images
\usepackage{lipsum} % Used for inserting dummy 'Lorem ipsum' text into the template
\usepackage{ amssymb }
\usepackage{slashed}
\usepackage{tikz}
\usepackage{comment}
\usepackage{natbib}
\usepackage[caption=false,listofformat=parens, subrefformat=parens]{subfig}
\usepackage{ marvosym }
\usepackage{array}
\usepackage{amsthm} % Theorem Formatting

\let\baraccent=\= % rename builtin command \= to \baraccent
\renewcommand{\=}[1]{\stackrel{#1}{=}} % for putting numbers above =

\theoremstyle{definition}

\theoremstyle{remark}

\newcolumntype{C}[1]{>{\centering\let\newline\\\arraybackslash\hspace{0pt}}m{#1}}

\begin{document}
\title{Attraction of indirect excitons in van der Waals heterostructures with three  semiconducting layers}

\author{M. Sammon} 
\email[Corresponding author: ]{sammo017@umn.edu} 
\affiliation{School of Physics and Astronomy, University of Minnesota, Minneapolis, MN 55455, USA}
\author{B. I. Shklovskii} 
\affiliation{School of Physics and Astronomy, University of Minnesota, Minneapolis, MN 55455, USA}

\received{\today}

\begin{abstract}
We study a capacitor made of three monolayers of transition metal dichalcogenide (TMD) separated by hexagonal Boron Nitride (hBN). We assume that the structure is symmetric with respect to the central layer plane. The symmetry includes the contacts: if the central layer is contacted by the negative electrode, both external layers are contacted by the positive one. As a result a strong enough  voltage $V$ induces electron-hole dipoles (indirect excitons) pointing towards one of the external layers. Antiparallel dipoles attract each other at large distances. Thus, the dipoles alternate in the central plane forming a 2D antiferroelectric with negative binding energy per dipole. The charging of a three-layer device is a first order transition, and we show that if $V_1$ is the critical voltage required to create a single electron-hole pair and charge this capacitor by $e$, the macroscopic charge $Q_c = eSn_c$ ($S$ is the device area) enters the three-layer capacitor at a smaller critical voltage $V_{c} < V_{1}$. In other words, the differential capacitance $C(V)$ is infinite at $V = V_{c}$. We also show that in a contact-less three-layer device, where the chemically different central layer has lower conduction and valence bands, optical excitation creates indirect excitons which attract each other, and therefore form antiferroelectric exciton droplets. Thus, the indirect exciton luminescence is red shifted compared to a two-layer device. 
\end{abstract}
\maketitle

In a standard parallel-plate capacitor, the capacitance $C$ is equal to the ``geometric capacitance" $C_g = \varepsilon S/4 \pi d$ (in Gaussian units), where $\varepsilon$ is the dielectric constant of the medium separating the two plates, $S$ is the area of each plate, and $d$ is the distance between them.  The expression $C = C_g$ is correct when both electrodes are made from a ``perfect" metal, which by definition screens the electric field with a vanishing screening radius. This condition fails if both sides of the capacitor are made of a layer of an intrinsic semiconductor and the applied voltage generates in them an equal small density $n$ of a two-dimensional electron (2DEG) and hole (2DHG) gas. For example, one can think about two separately contacted monolayers of intrinsic MoSe$_2$ separated by a few hexagonal boron nitride (hBN) layers with total width $d$. If in both the 2DEG and the 2DHG $na^2 \ll 1$, where $a = \varepsilon \hbar^2/me^2$ is the carrier Bohr radius and $m$ is its effective mass, electrons and holes created in opposite MoSe$_2$ layers can be treated as classical point like particles. It was shown~\cite{Brian2010} that if in addition $nd^2 \ll 1$, the capacitor charge $Q=enS$ grows with $V$ as $Q(V)  \propto (V-V_1)^{2/3}$, where
\begin{equation}\label{eq:V_1}
eV_1=E_g-E_{ex},
\end{equation}
  is the critical voltage required to create a single isolated electron-hole pair in an intrinsic semiconductor, $E_g$ is the bandgap of the semiconductor, and $E_{ex}$ is the binding energy of the electron-hole pair. The differential capacitance $C(n)\equiv dQ/dV$ becomes much larger than $C_g$ and grows as $0.37C_g/(nd^2)^{1/2}$ with decreasing $n$. As a function of $V$ the capacitance $C(V) \propto (V-V_1)^{-1/3}$.

\begin{figure}[htb]
	\includegraphics[width=\linewidth]{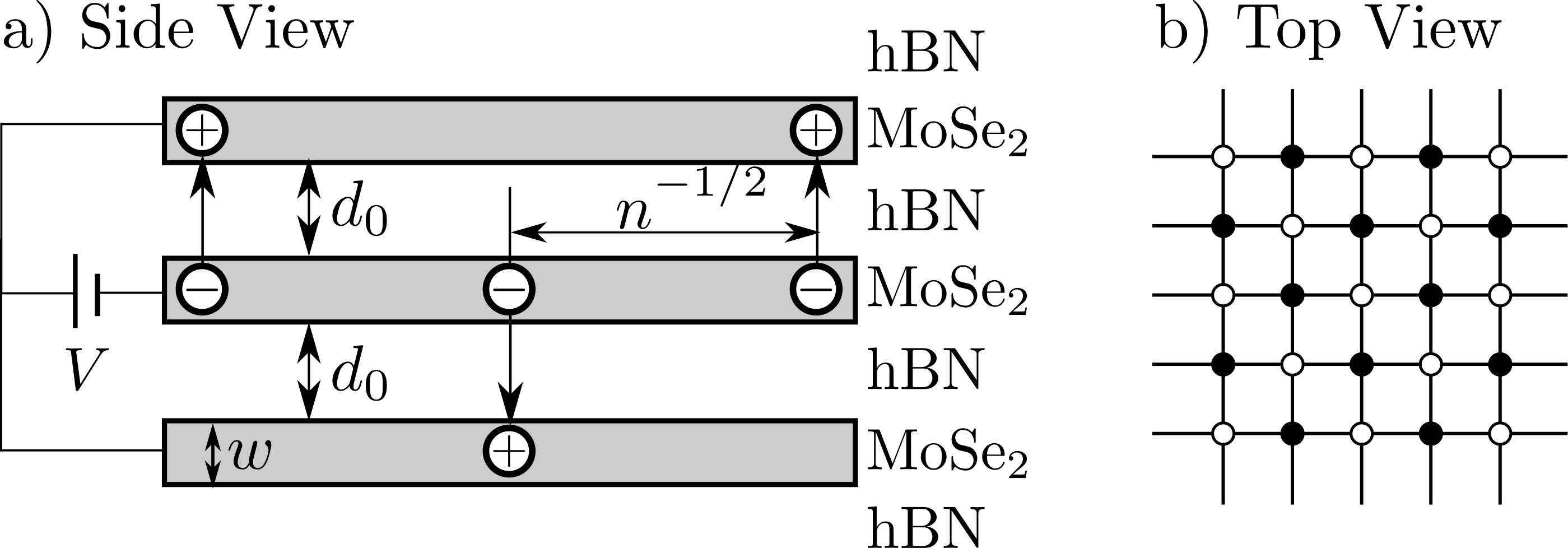}
	\caption{a) Cross section of a capacitor made of three MoSe$_2$ monolayers of width $w$ shown in gray.
		Each spacer of width $d_0$ has the same number of hBN layers, which are labeled. The outer top and bottom layers are covered by hBN as well.   
		The capacitor charges via the creation of alternating up and down pointing electron-hole dipoles (indirect excitons) shown by arrows. The electrons in the central plane form a square lattice with a lattice constant $n^{-1/2}$.
	b) Top view of the square lattice of alternating dipoles. White (black) circles correspond to dipoles whose orientation points up (down). Each orientation forms its own square sublattice.  } \label{fig:Threelayers}
\end{figure}

This anomalous capacitance growth near $V_1$ is due to the fact that each electron in the 2DEG is bound to a hole in the 2DHG of the other layer, forming an indirect exciton with a dipole moment $ed$.
 At $nd^2 \ll 1$, parallel dipoles are separated from each other by a large distance. Therefore, their repulsion is weak and provides a weak resistance to further capacitor charging leading to a diverging capacitance as $V\rightarrow V_1$ from above.  A similar anomalously large capacitance was predicted when one layer is replaced by a metallic plane.~\cite{Brian2010}  A capacitance 40\% larger than the geometrical value, which may be a result of this phenomena, was reported in YBCO/LAO/STO nanostructures.~\cite{Ashoori2011} A similar effect was predicted in graphene-metal capacitors placed in a strong perpendicular magnetic field which localizes carriers.~\cite{Brian2013} A capacitance that is $20\%$ larger than the geometrical one was observed in this case.~\cite{Kretinin2013}

The strong capacitance anomaly in the two-layer device is due to the discreteness of charge and their strong correlations at small densities $n$. In this paper we explore similar correlation physics in three-layer devices with a symmetry plane. For example, we may think about three monolayers of intrinsic MoSe$_2$ each of width $w$, separated by the same number of hBN layers of total width $d_0$ on each side, so that the structure is symmetric with respect to the central layer midplane. The symmetry is not only geometrical, but also includes the voltage contacts: if the central layer is contacted by the negative electrode, both external layers are contacted by the positive electrode. Correspondingly, an equal number of voltage induced indirect excitons, each with a dipole moment $ed=e(d_0+w)$, are directed from the central plane to the top and to the bottom
(see Fig.\ \ref{fig:Threelayers}a). At large distances along the plane, two antiparallel dipoles attract each other, while at distances smaller than $d$ they repel each other. It is natural to assume that as a result the dipoles form a two-dimensional antiferroelectric square lattice. This lattice is similar to NaCl, where Na-like and Cl-like sites are occupied by up and down pointing dipoles, i.e.\;all nearest neighbor dipoles are antiparallel (see Fig.\ \ref{fig:Threelayers}b) . Electrons of the central plane form a square lattice with the lattice constant $n^{-1/2}$.

\begin{figure}[t!]
	\includegraphics[width=\linewidth]{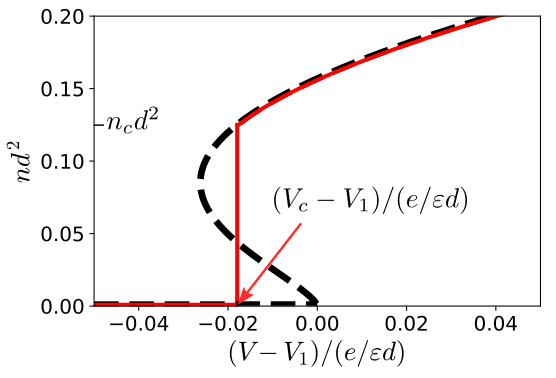}
	\caption{The dimensionless density $nd^2$ as a function of dimensionless voltage $(V-V_1)/(e/\varepsilon d)$  for the three-layer device shown in Fig.\,\ref{fig:Threelayers}. The dashed curve shows the curve $n(V)$ obtained from Eq.\,(\ref{eq:Voltage_def}), while the solid red curve shows the equilibrium $n(V)$ curve obtained using Maxwell's rule. We see that in equilibrium, the density jumps to a value $n_c$ at the critical voltage $V_c$. Thus the capacitor charge experiences a first order phase transition with growing $V$. } \label{fig:density_voltage}
\end{figure}

We show below that at low temperatures when the applied voltage $V$ grows, the attraction between indirect excitons in the three-layer device causes a first order phase transition (see Fig.\,\ref{fig:density_voltage}).
 While at small $V$ there are no dipoles and the capacitor remains uncharged, at some critical value $V=V_{c} < V_1$ the whole lattice of alternating dipoles emerges. This means that a macroscopic charge $Q_c=eSn_c$, where $n_c=0.13d^{-2}$ and $S$ is the device area, enters this capacitor. Thus, the differential capacitance $C$ has a $\delta$-peak at $V = V_{c}$.  At $V > V_c$, as  $n$ continues to grow the capacitance slowly approaches its normal geometric value $2C_g$. The giant 
$\delta$-peak of the capacitance at $V=V_{c}$ can be thought of as an enhanced version of the anomaly $C(V) \propto (V-V_1)^{-1/3}$ near $V=V_1$ predicted for a two-layer capacitor.\cite{Brian2010} A similar $\delta$-peak capacitance was predicted in a 3D nanocrystal film gated by an ionic liquid in which the ions penetrate between nanocrystals.~\cite{Tianran2011}

%Although above we talked about three-layer device made of MoSe$_2$ difficulties with making separate contacts to all three layers can make a similar graphene/hBN device more attractive. However, in this case we need a strong perpendicular magnetic field to localize electrons. 

For a quantitative description of the three-layer capacitor we assume the density is such that $na^2\ll1$ so that we may treat all charges classically. The differential capacitance of such a device can be determined from the total electrostatic energy $E$ of the system as 
\begin{equation}\label{eq:Cap_def}
C^{-1}=\frac{1}{(eS)^2}\frac{d^2E}{dn^2}.
\end{equation}
The energy $E$ of this system of classical charges can be written as 
\begin{equation}\label{eq:Energy_tot}
E=enSV_1+nSU ,
\end{equation}
where $V_1$ is the voltage necessary to create a single isolated electron-hole pair and is given by Eq.\,(\ref{eq:V_1}), while $U$ is the interaction energy per electron-hole pair in the system. We can further separate the interaction energy as $U=U_e+U_h$, where $U_e$ is the contribution to $U$ from the electrons interacting with all other charges, while $U_h$ is the contribution to $U$ from the holes interacting with all other charges. It should be clarified that in $U_e$ and $U_h$ we neglect the interaction between electrons and holes of the same pair. Assuming the electrons and holes can be treated classically, the binding energy $E_{ex}$ in Eq.\,(\ref{eq:V_1}) is given by $e^2/(\varepsilon d)$. Here and below we use the Coulomb potential with an effective dielectric constant $\varepsilon \simeq 5$ which is close to the dielectric constant of hBN. This effective medium potential was used in previous studies of electron-hole interactions in bilayer TMD heterostructures separated by several layers of hBN,\cite{Fogler2014} and is in contrast with the Rytova-Keldysh potential used for a single TMD layer in air. Its use for our system can be justified because the distance between neighboring dipoles $n_c^{-1/2}$  is much larger than the distance $2 w$ at which the electric field lines spread over the entire structure. Here the factor 2 comes from the ratio of the dielectric constant of MoSe$_2$ and hBN, and $w$ is the thickness of a monolayer of MoSe$_2$.

\begin{figure}
 	\includegraphics[width=\linewidth]{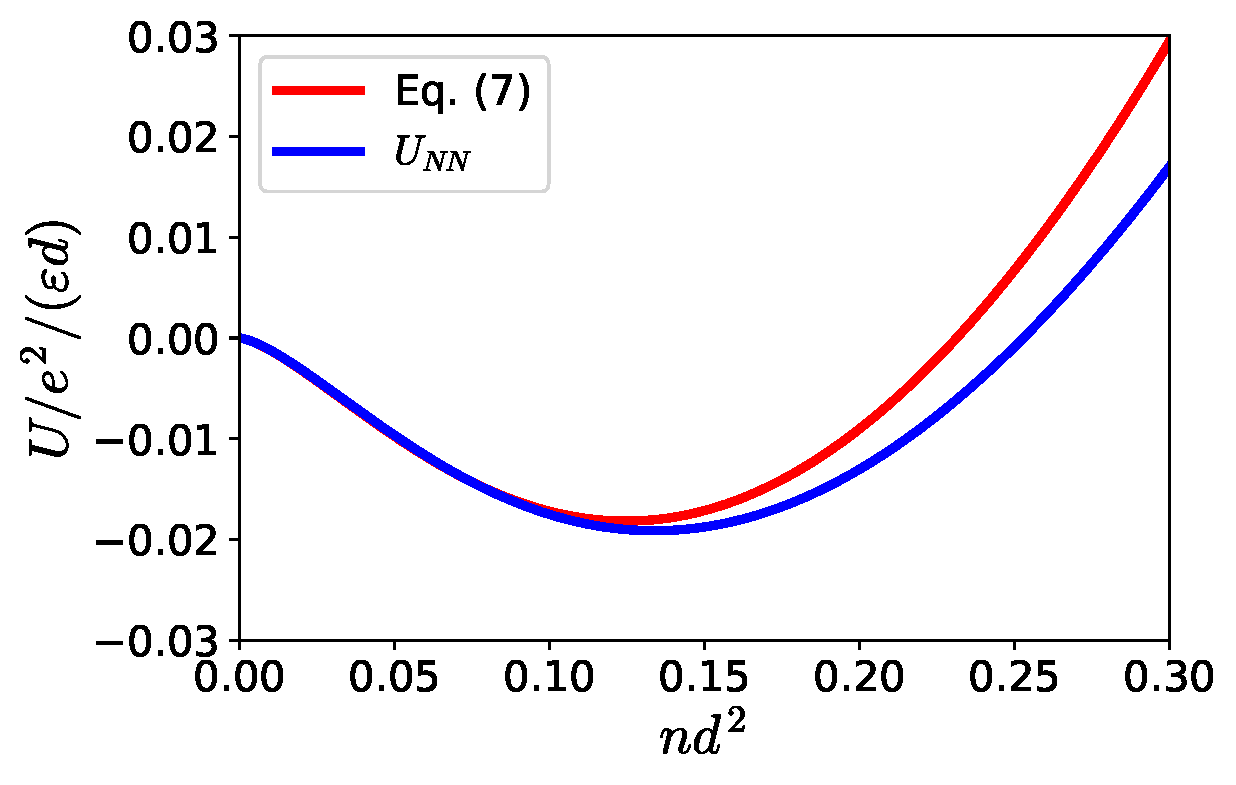}
 	\caption{Dimensionless interaction energy $U/e^2/(\varepsilon d)$ vs the density $nd^2$. The red curve is obtained from Eq.\,(\ref{eq:Energy_tot2}), while the blue curve labeled $U_{NN}$ is an approximation which only takes into account the nearest neighbor sites of each sublattice. }\label{fig:energy}
 \end{figure}

Each hole in an external plane pairs with an electron in the central plane in such a way that the orientation of the dipoles alternates between nearest neighbor sites of the electrons in the central plane square lattice, as shown in Fig.\,\ref{fig:Threelayers}(b). Let us consider the electron-hole pair located at the central white site in Fig.\,\ref{fig:Threelayers}(b). For the electron at the origin, we can write
\begin{equation}\label{eq:Energy_central}
U_e=\frac{1}{2}\sum_{\alpha\neq0}\left(\frac{e^2}{\varepsilon r_{\alpha}}-\frac{e^2}{\varepsilon\sqrt{r_{\alpha}^2+d^2}}\right),
\end{equation}
where $\alpha$ is an index labeling the electron lattice sites, $\alpha=0$ is defined as the origin, and $r_{\alpha}$ is the distance between site $\alpha$ and the origin. The factor $1/2$ accounts for the double counting when computing the interaction energy $U$.  For the hole that is also located at the origin, we can use the fact that the electron and hole form a dipole with a particular orientation (in this case upwards) to separate $U_h$ as $U_h=U_{h1}+U_{h2}$. Here
\begin{equation}\label{eq:Energy_ext1}
U_{h1}=\frac{1}{2}\sideset{}{^{\circ}}\sum_{\alpha\neq0}\left(\frac{e^2}{\varepsilon r_{\alpha}} -\frac{e^2}{\varepsilon\sqrt{r_{\alpha}^2+d^2}}\right),
\end{equation}
is the contribution from the interaction of the hole with dipoles with the same orientation as the origin dipole (white sites), while 
\begin{equation}\label{eq:Energy_ext2}
U_{h2}=\frac{1}{2}\sideset{}{^{\bullet}}\sum_{\alpha\neq 0}\left(\frac{e^2}{\varepsilon\sqrt{r_{\alpha}^2+4d^2}} -\frac{e^2}{\varepsilon\sqrt{r_{\alpha}^2+d^2}}\right),
\end{equation}
is the contribution from the interaction of the hole with dipoles of the opposite orientation(black sites). The symbols next to the summation indicate that the sums are restricted to the corresponding sublattice shown in Fig.\,\ref{fig:Threelayers}(b).
Upon inspection, it is clear that Eqs. (\ref{eq:Energy_central}) and (\ref{eq:Energy_ext1}) are similar so that we can write $U$ as 
\begin{equation}\label{eq:Energy_tot2}
U=\frac{e^2n^{1/2}}{2\varepsilon}\left(g(nd^2)+\frac{1}{\sqrt{2}}g(nd^2/2)+h(nd^2)\right),
\end{equation}
where  
\begin{equation}\label{eq:g(x)}
g(x)=4\sum_{i=1}^{\infty}\sum_{j=0}^{\infty}\left(\frac{1}{\sqrt{i^2+j^2}}-\frac{1}{\sqrt{i^2+j^2+x}}\right),
\end{equation}
and 
\begin{equation}\label{eq:h(x)}
h(x)=4\sum_{i=1}^{\infty}\sideset{}{^{\bullet}}\sum_{j=0}^{\infty}\left(\frac{1}{\sqrt{i^2+j^2+4x}}-\frac{1}{\sqrt{i^2+j^2+x}}\right),
\end{equation}
and we have rewritten the site index $\alpha$ using the integers $i$ and $j$ of the electron lattice coordinates in units of $n^{-1/2}$. For the summation over the black sublattice in Eq.\,(\ref{eq:h(x)}), we restrict ourselves to values of $i$ and $j$ such that $i+j$ is odd. Both summations are convergent for any $x$. The results of this summation are shown by the red curve in Fig.\,\ref{fig:energy} as a plot of $U/e^2/(\varepsilon d)$ vs $nd^2$. We see that the interaction energy is negative for a finite range of densities due to the attraction between nearest neighbor dipoles with opposite orientation. At $n_cd^2=0.13$ it reaches a minimum value of $U=-0.018e^2/\varepsilon d$. In order to better understand this, we compare this result to the energy obtained from only the nearest neighbor sites of each sublattice, shown by the labeled $U_{NN}$ curve (blue curve) in Fig.\,\ref{fig:energy}. We see that for small $nd^2$ the energy is almost completely determined by these nearest neighbors, with significant deviation only appearing beyond the minimum of $U$.

Once the energy $E$ is found, we can find the voltage as 
\begin{equation}\label{eq:Voltage_def}
V=V_1+\frac{1}{e}\frac{d(nU)}{dn}.
\end{equation}

\begin{figure}[t]
 	\includegraphics[width=\linewidth]{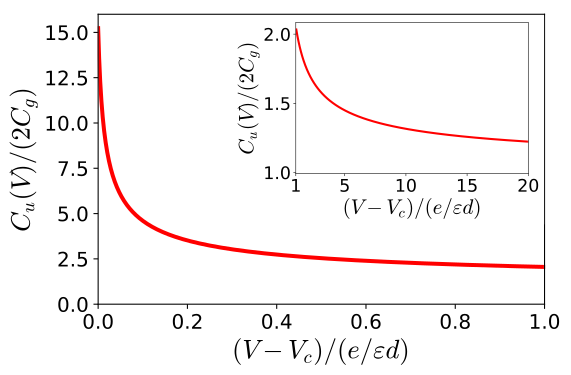}
 	\caption{The dimensionless capacitance $C_u(V)/(2C_g)$ as a function of the dimensionless voltage $(V-V_c)/(e/\varepsilon d)$ corresponding to the upper branch of the red (solid) curve shown in Fig.\,\ref{fig:density_voltage}. The inset shows $C_u(V)/(2C_g)$ over a 19 times larger range of $(V-V_c)/(e/\varepsilon d)$.} \label{fig:capacitance_voltage}
 \end{figure}

Our main result is shown in Fig.\,\ref{fig:density_voltage} as a plot of the density $nd^2$ as a function of the voltage $(V-V_1)/(e/\varepsilon d)$. The dashed curve is obtained from Eq.\,(\ref{eq:Voltage_def}). Most noticeable is that there is a range in which there are three densities for each voltage: a lower branch along $n=0$, a middle branch, and an upper branch. Within the middle branch, the capacitance defined by Eq.\,(\ref{eq:Cap_def}) is negative and this region is thermodynamically unstable and is inaccessible. Thus in experiment, we do not expect the density to change continuously along the dashed curve, but instead along the curve shown in red where the density jumps to a value 
\begin{equation}\label{eq:n_c}
n_c=0.13d^{-2},
\end{equation}
 at a critical voltage 
\begin{equation}\label{eq:V_c}
V_c=V_1-0.018\frac{e}{\varepsilon d}.
\end{equation} 
Here $V_c$ is determined by Maxwell area rule\cite{landau_statphys1}
\begin{equation}\label{eq:equilibrium}
\int_0^{n_c} n(V)dV=0,
\end{equation}
where the integral is taken along the dashed curve in Fig.\,\ref{fig:density_voltage}. At $V=V_c$ the two regions lying between the vertical red line and the dashed curve have equal area. This rule is well known for the van der Waals liquid-gas pressure-volume isotherm.
\footnote{In this analogy, $n$ plays the role of volume while the voltage $V$ plays the role of pressure.}   
 It is worth noting that $n_c$ obtained from Maxwell's area rule is the same $n_c$ at which $U$ reaches its minimum value. 
 As the density abruptly jumps, there is a $\delta$-peak in the capacitance at $V=V_c$. For $V\geq V_c$ we can write the capacitance as
 \begin{equation}\label{eq:C(V)}
 C(V)=eSn_c\delta(V-V_c)+C_u(V),
 \end{equation} 
 where the non-singular capacitance $C_u(V)$ is obtained by differentiating the upper branch of the $n(V)$ curve shown in Fig.\,\ref{fig:density_voltage} with respect to $V$ and is shown in Fig.\,\ref{fig:capacitance_voltage}. As $V$ approaches $V_c$ from above, $C_u(V)$ grows as $(V-V_c+0.01e/\varepsilon d)^{-1/2}$, and attains a very large maximum value $C_u(V_c)\simeq30 C_g$, where $C_g$ is the geometrical capacitance of the capacitor formed by either the central and upper planes or the central and lower planes. At larger voltages $V\gg e/\varepsilon d$ it approaches $2C_g$ corresponding to the geometric value of the three-layer system as shown in the inset of Fig.\,\ref{fig:capacitance_voltage}.
 
So far we have been dealing with very low temperatures and have ignored disorder. Temperature and disorder smear the $\delta$-function as well as the low voltage peak of the non-singular large voltage tail of $C_u(V)$. When the width of the $\delta$-function reaches $V_1 - V_c$, the two peaks in the capacitance merge to form a single peak. Because this happens at $C_u \simeq 30 C_g$, a very large peak of the capacitance (much larger than in the case of two layers) survives in the presence of disorder or higher temperatures. It is easy to imagine that the  measured capacitance peak is 5-10 times larger than the geometrical value. The reason for the early merging of the $\delta$-function with the non-singular peak is that the optimal distance between electrons in the central plane $\sim n_c^{-1/2} \simeq 3d$ is relatively large and makes both the optimal energy and the voltage scale $(V_1-V_c)$ of the dipole configuration in Fig.\,\ref{fig:density_voltage} relatively small. We can estimate the scale of temperature at which thermal fluctuations destroy the effect from the minimum in the interaction energy $U=-0.018e^2/\varepsilon d$ shown in Fig.\,\ref{fig:energy}. For $\varepsilon=5$ and $d\simeq 1$ nm for a three layer thick hBN spacer, we find at $T\simeq60$ K thermal fluctuations begin to dominate.

We have also ignored quantum effects. Typically the localization length  $\xi$ of electrons in the central plane
can be comparable with $d$, so that quantum effects may modify the energy of the three-layer system at large enough $n$ even at zero
temperature and disorder.~\cite{Fogler2014} However, even in such a case, at small $n$ the energy of 
the electron-hole dipoles (excitons) is dominated by their dipole-dipole attraction and charging occurs by 
the first order transition. Quantum mechanics can still somewhat reduce $n_c$  and $(V_1-V_c)$.
Quantum Monte-Carlo simulations similar to those in Refs.\,[\onlinecite{Palo(2002),Schleede2012,Needs2013}] 
are necessary to address these changes quantitatively.

\begin{figure}[t]
	\includegraphics[width=\linewidth]{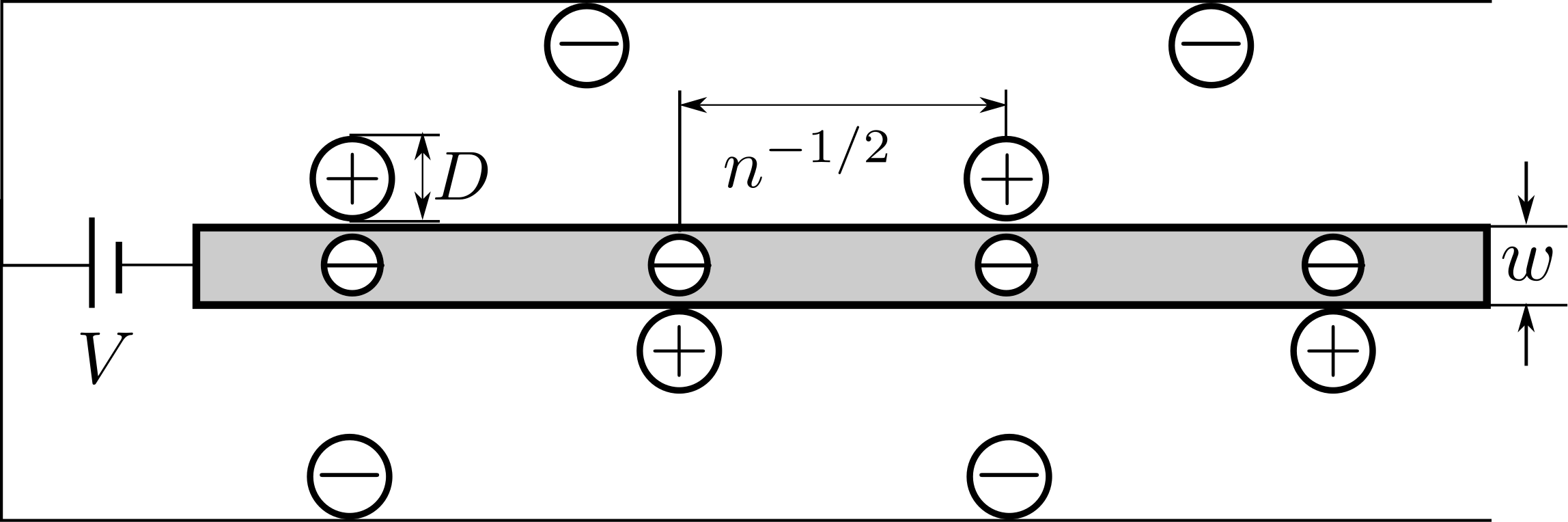}
	\caption{Schematic drawing of a TMD monolayer gated on both sides by an ionic liquid. A positive electrode immersed in the ionic liquid forces a concentration $n$ of excessive positive ions to the surface of the TMD monolayer, while simultaneously attracting an equal concentration of excessive negative ions to the electrode surface (background ions of the net neutral ionic liquid are not shown). Each excessive positive ion binds an electron in the TMD, forming a dipole with arm length $(D+w)/2$. Similar to Fig.\,\ref{fig:Threelayers}, the oppositely oriented dipoles attract each other and the electrons arrange in a square lattice of lattice constant $n^{-1/2}$.}\label{fig:ionicliquid}
\end{figure}

\begin{figure}[t!]
	\includegraphics[width=\linewidth]{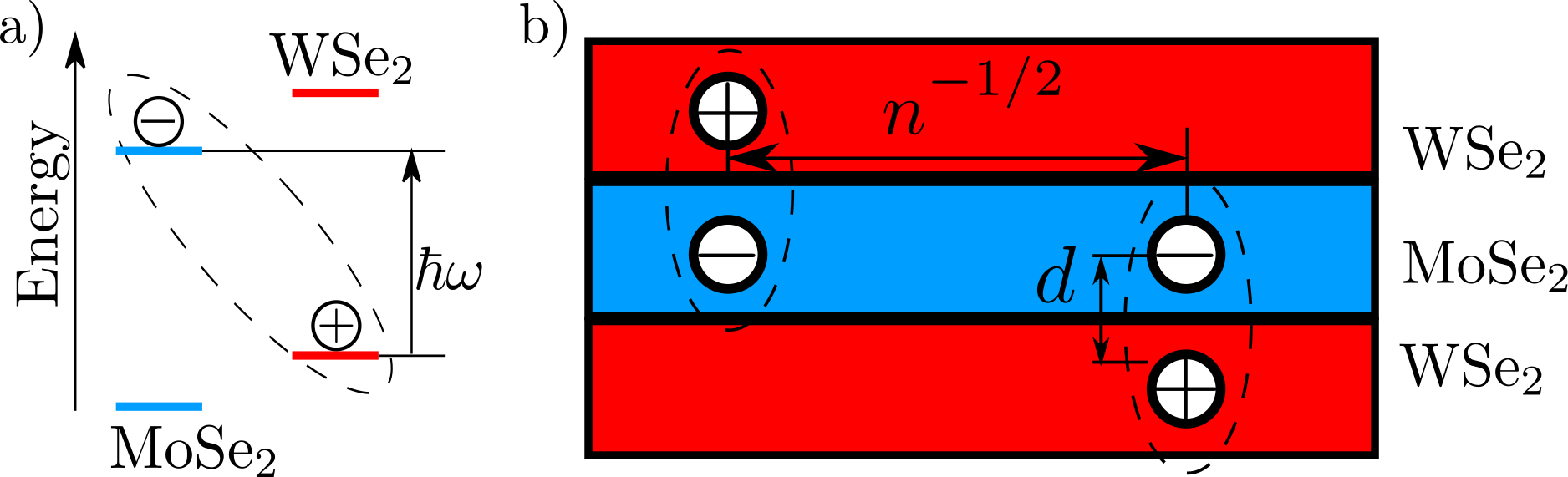}
	\caption{a) Band alignment in MoSe$_2$/WSe$_2$ bilayer. b) Schematic of a trilayer WSe$_2$/MoSe$_2$/WSe$_2$ device for optical studies of spatially interacting indirect excitons. When the device is illuminated at low temperatures, the type II band alignment of neighboring WSe$_2$/MoSe$_2$ monolayers (see inset) allows the formation of indirect excitons consisting of an electron in MoSe$_2$ and a hole in WSe$_2$. Excitons of opposite polarity attract each other and form a crystal with alternating dipoles with concentration $n_c$ shown in Fig.\,\ref{fig:Threelayers}. }\label{fig:optical}
\end{figure}

Devices similar to those shown in Figs.\,\ref{fig:Threelayers} and \ref{fig:ionicliquid} can be also made from graphene monolayers, however in this case the classical model leading to attraction between indirect excitons becomes useful only in strong magnetic fields such that  $n l_B^2 = n\hbar c/eB\ll1$,
where $l_B$ is the magnetic length. In devices with $d \ll l_B$ this condition may substantially reduce $n_c$. 

Three-layer devices made of MoSe$_2$ can face difficulties in making separate contacts to all 
three layers. Therefore, a similar device made of a single MoSe$_2$ monolayer gated from both sides by an ionic liquid can be more attractive. 
In this case the dipoles are formed by electrons of the MoSe$_2$ monolayer bound to excessive
positive ions, which stick 
to the monolayer in alternating positions above and below it (see Fig.\,\ref{fig:ionicliquid}). At small electron densities such a device is quantitatively similar to the three layer device described with a dipole moment $e(D+w)/2$, where $D$ is the ionic diameter.

Above we talked about the capacitance of three-layer devices. Three-layer devices can be also used for optical studies of spatially indirect interacting excitons. It has been shown\cite{Rivera(2015),Geim(2019)} that in bilayer MoSe$_2$/WSe$_2$ structures, the type II band alignment of the MoSe$_2$ and WSe$_2$ monolayers allows the formation of indirect excitons, in which an electron in MoSe$_2$ binds to a hole in WSe$_2$ (see Fig.\ref{fig:optical}a). 
 Because of the weak overlap of the electron and hole wavefunctions, these excitons decay slowly enough to form the ground state which minimizes their repulsion. In the photoluminescence experiments on the MoSe$_2$/WSe$_2$ device of Ref.\,[\onlinecite{Rivera(2015)}], it was observed that the indirect exciton luminescence line blueshifts as the intensity of the laser increases due to the dipole-dipole repulsion of the indirect excitons. In a trilayer device,\cite{Plochocka(2017),Choi(2018)} such as WSe$_2$/MoSe$_2$/WSe$_2$ (and similar devices with symmetric hBN spacers), we instead predict an attractive interaction between indirect excitons formed from opposite WSe$_2$ layers (see Fig.\,\ref{fig:optical}b). At low illumination intensities these excitons condense into droplets of density $n_c$ which do not interact with each other. These droplets are different from the exciton droplets in 3D semiconductors proposed by Keldysh and Kozlov.\cite{Keldysh(1968)} Those droplets are formed by the van der Waals attraction between excitons and occurs when the excitons are at distances of order $a$. Our droplets are the result of the electrostatic dipole-dipole interaction and the excitons are separated by the larger than $a$ distance $n_c^{-1/2}$ set by our classical theory.  In such a device, the luminescence peak should be redshifted.
\footnote{We can extrapolate the classical energy of a crystal of alternating dipoles to estimate the redshift as $0.02e^2/(\varepsilon d)\sim 6$ meV, where we have used $\varepsilon\simeq 7$ and $d=7$ $\AA$ for monolayer TMDs.\cite{Laturia(2018)}}
  The luminescence line of excitons in these droplets should not change with the laser intensity until the intensity becomes so large that the droplets fill the entire sample.

We are grateful to L.\,V. Butov, M.\,M. Fogler,  Q. Shi, and B. Skinner for useful discussions. M.\,Sammon was supported primarily by the NSF through the University of Minnesota MRSEC under Award No. DMR-1420013.

%Things to add:
%disorder
%temperature
%graphene strong magnetic field  
%excitons
%optics
%quantum mechanics

%\vspace{200mm}

%\bibliography{GiantRef} 

%merlin.mbs apsrev4-1.bst 2010-07-25 4.21a (PWD, AO, DPC) hacked
%Control: key (0)
%Control: author (72) initials jnrlst
%Control: editor formatted (1) identically to author
%Control: production of article title (-1) disabled
%Control: page (0) single
%Control: year (1) truncated
%Control: production of eprint (0) enabled
%

\end{document}